# Investigating the impact of kernel harmonization and deformable registration on inspiratory and expiratory chest CT images for people with COPD

Aravind R. Krishnan[a], Yihao Liu[a], Kaiwen Xu[b], Michael E. Kim[c], Lucas W. Remedios[c], Gaurav Rudravaram[a], Adam M. Saunders[a], Bradley W. Richmond[g,h], Kim L. Sandler[i], Fabien Maldonado[j,k], Bennett A. Landman[a,c,d,e,f], and Lianrui Zuo[a]

[a]Department of Electrical and Computer Engineering, Vanderbilt University, Nashville, TN, USA,
[b]Insitro, South San Francisco, CA, USA,
[c]Department of Computer Science, Vanderbilt University, Nashville, TN, USA, [d]Department of Biomedical Engineering, Vanderbilt University, Nashville, TN, USA, [e]Department of Radiology and Radiological Sciences, Vanderbilt University Medical Center, Nashville, TN, USA, [f]Vanderbilt University Institute of Imaging Science, Vanderbilt University Medical Center, Nashville, TN, USA,
[g]Department of Allergy, Pulmonary and Critical Care Medicine, Vanderbilt University Medical Center, Nashville, TN, USA, [h]Department of Veterans Affairs Medical Center, Nashville, TN, USA, [i]Department of Radiology, Vanderbilt University Medical Center, Nashville, TN, USA, [j]Department of Medicine, Vanderbilt University Medical Center, Nashville, TN, USA, [k]Department of Radiology, Vanderbilt University Medical Center, Nashville, TN, USA

## ABSTRACT

Paired inspiratory-expiratory computed tomography (CT) scans enable quantification of gas trapping alterations due to small airway disease and emphysema through the motion of the lung tissue for people with chronic obstructive pulmonary disease (COPD). Deformable image registration of these paired CT scans is often used to assess the regional volumetric changes in the lung. However, variations in reconstruction protocols, particularly the reconstruction kernels between paired inspiratory-expiratory scans are often overlooked, and these variations introduce errors during quantitative image analysis. In this work, we propose a two-stage pipeline to harmonize reconstruction kernels between paired inspiratory-expiratory scans and perform deformable image registration for data acquired from the COPDGene study. We use a cycle generative adversarial network (GAN) for image synthesis to harmonize inspiratory scans reconstructed with a hard kernel (BONE) to match expiratory scans reconstructed with a soft kernel (STANDARD). We then perform deformable image registration to register the expiratory scans to the inspiratory scans. We validated harmonization by measuring emphysema using a publicly available segmentation algorithm, both before and after harmonization. Our results show that harmonization significantly reduces inconsistencies in emphysema measurement, decreasing the median emphysema scores from 10.479% to 3.039% with a reference median score of 1.305% from the STANDARD kernel as a harmonization target. We validate the registration accuracy by observing the Dice overlap between emphysema regions on the inspiratory, expiratory and deformed images. The Dice coefficient between the fixed inspiratory emphysema masks and deformably registered emphysema masks increases across different stages of registration with statistical significance ($p<0.001$). Additionally, we show that deformable registration is robust to kernel variation.

Corresponding author email: aravind.r.krishnan@vanderbilt.edu

## 1. INTRODUCTION

Chronic obstructive pulmonary disease (COPD) is characterized by irreversible airflow limitation due to inhalation of toxic gases or particles, resulting in changes in the lung parenchyma, bronchi, and pulmonary vessels[1]. Computed tomography (CT) scans obtained at inspiration and expiration present opportunities to quantitatively evaluate emphysema and air trapping using density-based thresholds either on individual or co-registered scans[2,3]. However, quantitative measurements to assess lung density are impacted by the CT reconstruction kernel[4]. A hard kernel has higher spatial resolution accompanied by noise while a soft kernel has lower spatial resolution and reduced noise[5]. The amount of noise introduced during reconstruction impacts the lung parenchyma and surrounding anatomy, that can lead to differences in measurements during quantitative imaging (**Figure 1**). Therefore, an appropriate choice of reconstruction kernel is necessary to ensure accurate quantitative imaging measures. Kernel harmonization facilitates accurate measurements for quantitative imaging[6]. Existing methods for kernel harmonization have investigated harmonization in paired reconstruction kernels and unpaired reconstruction kernels. In paired reconstruction kernels, a subject has paired reconstructions obtained from the same manufacturer, reconstructed using different kernels such that a one-to-one pixel correspondence exists between them. On the contrary, unpaired reconstruction kernels pose additional challenges in the form of difference in the scanner protocol and lack of one-to-one mapping between different subjects.

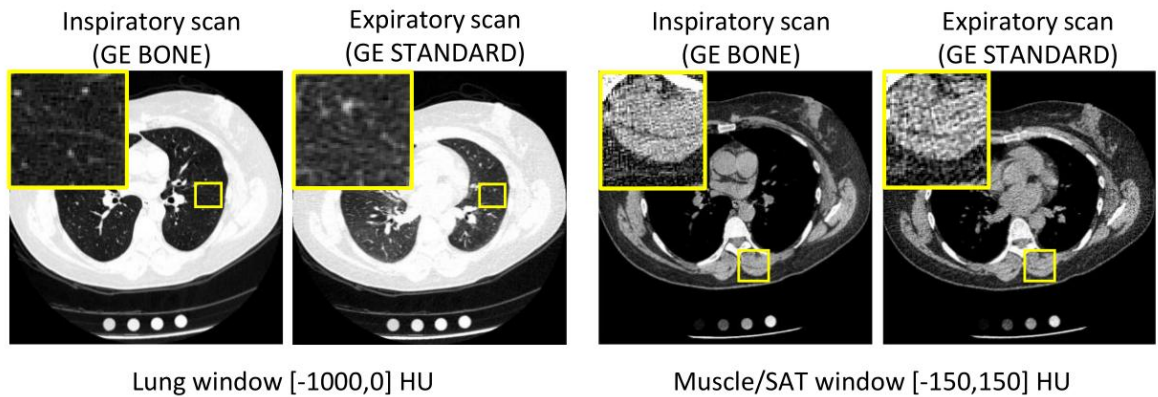

**Figure 1.** Paired inspiratory-expiratory scans present opportunities to quantify lung density measurements that include percent emphysema score and air trapping. However, variability in reconstruction kernels on paired inspiratory-expiratory scans for subjects having chronic obstructive pulmonary disease presents challenges in accurately quantifying lung density measurements. A BONE kernel has high spatial resolution accompanied by noise while a STANDARD kernel has lower spatial resolution and lesser noise compared to a BONE kernel. Harmonization of an inspiratory BONE kernel to an expiratory STANDARD kernel enforces consistency in lung density measurements, enabling accurate quantitative image analysis.

Recent advances in kernel harmonization have explored physics-based approaches and deep learning approaches. Saman Sotoudeh-Paima et al.[7] implemented a physics-based harmonization method to standardize images to a reference iso resolution and noise conditions by incorporating the modulation transfer function (MTF) and global noise index in their harmonizer model, successfully eliminating differences between reconstruction kernels for emphysema biomarkers for people with COPD. Similarly, Zarei et al.[8] implemented a physics-based neural network that successfully minimized inconsistencies in emphysema measurements in COPD cases. Deep learning methods on the other hand use convolutional neural networks (CNN) and image-to-image translation for kernel harmonization of paired and unpaired reconstruction kernels through supervised and unsupervised approaches. Lee et al.[9] implemented a CNN model inspired by super resolution for kernel harmonization on paired kernels and showed that the CNN was able to mitigate differences in emphysema quantification across reconstruction kernels through residual learning. Choe et al.[10] implemented a CNN model for harmonizing radiomic features computed across different reconstruction kernels and showed that differences in the reconstruction kernels were mitigated, resulting in a high correlation between the radiomic features of the harmonized kernel and ground truth soft kernel. Our group[11] and Tanabe et al.[12] implemented a pix2pix approach for paired CT kernel harmonization, successfully minimizing differences in emphysema quantification, muscle and subcutaneous adipose tissue measurements. Our group[6] extended the paired kernel harmonization approach to unpaired reconstruction kernel by developing a multipath cycle GAN that uses a shared latent space. The shared latent space was able to minimize differences in emphysema quantification between paired and unpaired kernels.

With the availability of paired inspiratory-expiratory scans, image registration maps voxels from the inspiratory scan to the corresponding expiratory scan[13]. Through registration, local change in air volume can be measured based on the change in lung density. Functional change in the lung volume can be quantified by calculating the Jacobian determinant on the warp field, reflecting biomechanical properties of the tissue. Additionally, registration allows Parametric Response Mapping (PRM) of paired inspiratory-expiratory scans, that benefits COPD phenotyping[14]. Recent advances in assessing functional changes in lung volume in COPD patients have provided valuable insights. Bhatt et al.[15] co-registered inspiratory-expiratory scans from the COPGene study and computed the Jacobian determinant to assess changes in emphysematous and regular lung tissue. Cohen et al.[16] studied the relationship between the Jacobian determinant and intensity-based metric of lung function with spirometry measures in COPD patients, showcasing positive correlations between the two. Koyama et al.[17] assessed three-dimensional lung motion in inspiratory expiratory scans to study pulmonary functional loss in smokers and destruction of lung and air trapping. Nishio et al.[18] developed a method to quantify extent of emphysema and small airways disease in COPD subjects using paired inspiratory-expiratory scans and deformable image registration. Hasenstab et. al[19] developed a convolutional network model for deformable lung registration on inspiratory-expiratory scans where the Jacobian determinant was incorporated in the objective function of the model to facilitate accurate air trapping in COPD subjects. While previous works explore harmonization of reconstruction kernels and registration of inspiratory-expiratory scans independently, the extent to which harmonization and image registration can impact studies involving paired inspiratory-expiratory chest CT scans is yet to be explored.

In this work, we first investigate the impact of kernel variations on quantitative measures and deformable image registration on paired inspiratory-expiratory scans for people with COPD. We then propose a two-stage pipeline which first harmonizes the inspiratory hard kernel images to the expiratory soft kernel images using unpaired image-to-image translation. We perform deformable image registration between inspiratory and expiratory scans with matched harmonized kernels (**Figure 2**). In order to assess the accuracy of both registration before and after harmonization, we apply the calculated deformation field to warp the emphysema mask of the moving expiratory image to the emphysema mask of the fixed inspiratory image and compare Dice overlap between the fixed and warped emphysema masks.

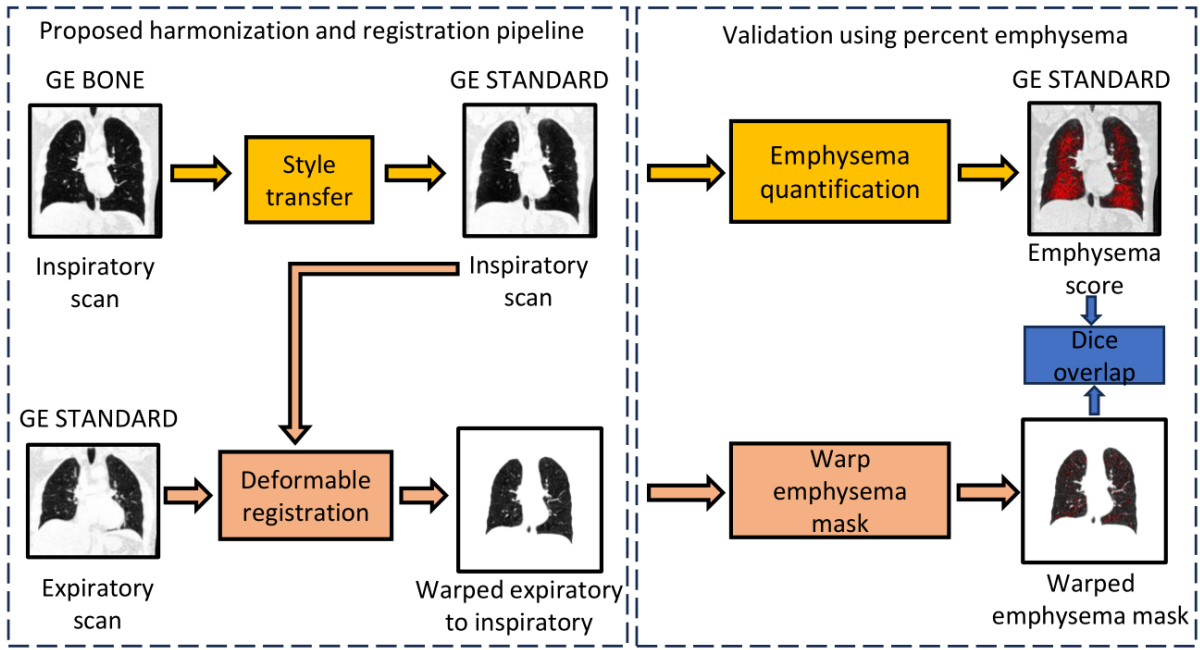

**Figure 2.** Differences in kernels can be mitigated through kernel harmonization, formulated as a style transfer problem. We perform unpaired kernel harmonization to harmonize a hard kernel (BONE) inspiratory scan to a soft kernel (STANDARD) scan and validate harmonization using percent emphysema quantification. Using the harmonized images, we perform deformable image registration of the expiratory CT to the inspiratory CT, leading to a deformed expiratory CT image. The warp field obtained from registration is applied to the emphysema mask obtained on the moving image and Dice overlap is computed between the fixed image emphysema mask and the deformably registered emphysema mask.

## 2. METHODOLOGY

We use data from the COPDGene study, a multicenter study that was carried out to identify genetic factors associated with COPD and characterize chest CT phenotypes that include gas trapping, airway wall thickening and emphysema[20]. A total of 10,000 participants were included in the COPDGene study. These participants were either non-Hispanic whites or African-Americans ranging between the ages of 45 and 80 with a minimum smoking history of ten pack years (excluding non-smoking controls). Spirometry analysis was conducted where the ratio of forced expiratory volume at one second

(FEV$_1$) and forced vital capacity (FVC) was computed. The Global Initiative for Chronic Obstructive Lung Disease (GOLD) system for classifying airflow obstruction in COPD was used to classify subjects into the following categories: GOLD unclassified, GOLD 0 (control), GOLD 1 (mild COPD), GOLD 2 (moderate COPD), GOLD 3 (severe COPD), and GOLD 4 (very severe COPD). Additionally, each subject had a pair of inspiratory and expiratory breath hold CT scans, reconstructed using hard and soft reconstruction kernels within a given vendor. We choose scans such that a subject has a pair of inspiratory-expiratory scans, reconstructed using different kernels. We select a total of 160 scans from the GE vendor, with 80 inspiratory scans reconstructed using the BONE (hard) kernel and 80 expiratory scans using the STANDARD (soft) kernel comprising of controls, moderate and severe COPD subjects. The slice thickness of the scans is 0.625 mm and the peak kilovoltage (kVp) is 120. The images were isotropic in the in-plane and anisotropic in the through-plane. We choose 50 scans each (25 controls, 25 cases) from the BONE and STANDARD kernel for the purpose of training image harmonization. The trained harmonization model is used to harmonize the remaining 30 (17 controls, 13 cases) scans that is used for image registration.

## 2.1 Data preprocessing

We convert the raw DICOM data to NIfTI using the dcm2niix tool[21] (version 1.0.2). During kernel harmonization, to ensure consistency in the range of Hounsfield units (HU) the images are clipped to [-1024, 3072] (HU) and then normalized to [-1,1] before training the model. We generate lung masks using a publicly available segmentation algorithm[22] and mask out the lungs by multiplying the lung masks with the corresponding CT volumes. All masked CT volumes are clipped between [-1024, 0] HU.

## 2.2 Harmonization using cycle GAN

Following our previous work[11], we formulate kernel harmonization as a style transfer problem and perform kernel harmonization between the inspiratory BONE kernels and expiratory STANDARD kernels using a cycle GAN[23]. The cycle GAN consists of a forward path where the source generator maps the real inspiratory BONE kernel image to a synthetic expiratory STANDARD kernel image and a backward path, where the target generator maps the real target expiratory STANDARD kernel image to a synthetic inspiratory BONE kernel image (**Figure 3**). The discriminators are responsible for classifying whether the generated images are real or synthetic. The generator model is a ResNet[24] consisting of 9 residual blocks. The discriminator is a PatchGAN[25] classifier that classifies $70 \times 70$ patches of images as real or synthetic. Instance normalization is used in the generator and discriminator.

We train the cycle GAN on two-dimensional gray scale axial isotropic slices of resolution $512 \times 512$ pixels. We breakdown the volumes into individual axial slices and obtain 25414 slices for the source domain and 24265 slices for the target domain. We train the cycle GAN model in parallel on two NVIDIA RTX A6000 GPUs for 50 epochs with a batch size of 12, using the Adam[26] optimizer and a learning rate of 0.0002. We implement a linear decay on the learning rate where the learning rate remains constant for the first 25 epochs and decays linearly for the next 25 epochs. The generator and discriminator are governed by an adversarial loss which is implemented using the least squares GAN (LSGAN) loss[27] and an L1 cycle consistency loss, with the latter weighted by 10 to ensure consistency in image contents during style transfer. We use the fifth epoch to generate the harmonized inspiratory scans based on qualitative assessment and compute emphysema scores by assessing the percentage of voxels lesser than -950 HU using lung masks obtained on the inspiratory, expiratory and harmonized inspiratory scans. The threshold of -950 HU has been shown to identify the microscopic and macroscopic extent of emphysema[28].

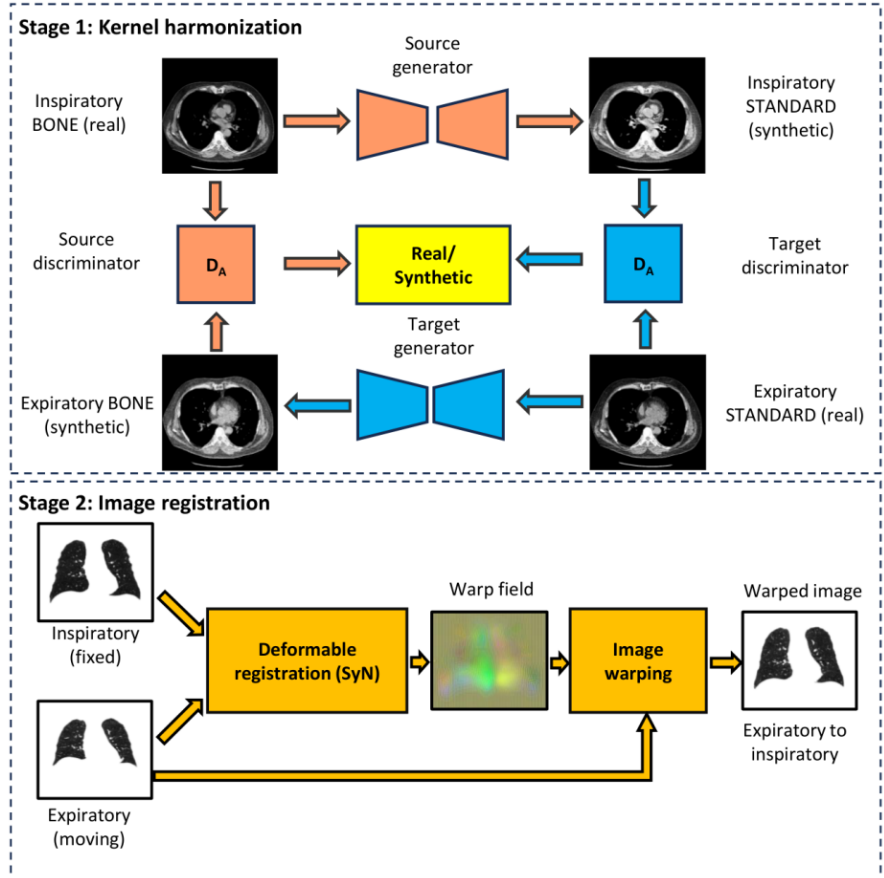

**Figure 3**. We use a cycle GAN to harmonize the inspiratory hard kernel to a synthetic inspiratory soft kernel image in the forward path while a synthetic expiratory hard kernel image is generated in the backward path. The discriminators govern the adversarial training and classify whether the images are real or synthetic. We register the harmonized and non-harmonized images using a combination of rigid, affine and deformable image registration. The registration produces a warp field that is applied to the moving image to obtain the deformed expiratory to inspiratory image.

**2.3 Deformable image registration**

We perform 3D deformable image registration at full resolution between the expiratory scan and inspiratory scans using the Advanced Normalization Tools (ANTs) package (https://github.com/ANTsX/ANTs) before and after harmonization. We use symmetric normalization[29] (SyN) as the registration method. We treat the expiratory scan as the moving image and the inspiratory scan as the fixed image and deformably register the images. We obtain the forward transform, inverse transform, image warped to the fixed space and the inverse warped image. Using the transforms, we deformably register the emphysema masks of the moving image to the fixed image for harmonized and non-harmonized images to study the overlap between the inspiratory and the warped expiratory to inspiratory emphysema mask. To assess improvement in registration, we perform a rigid registration between the moving expiratory image and the fixed inspiratory image followed by deformable registration at half the image resolution between the inspiratory and expiratory images. For all three stages of registration, Dice overlap is computed to study improvement.

## 3. RESULTS

We evaluate the efficacy of harmonization and registration independently followed by a joint analysis of the two approaches. Before harmonization, inspiratory scans reconstructed with the BONE kernel exhibit sharper structures as observed in the lung region while expiratory scans reconstructed with the STANDARD kernel are smoother compared to the inspiratory scan. After harmonization, inspiratory scans display smoothened lung regions similar to the expiratory scans (**Figure 4**). We qualitatively evaluate registration using a 10 × 10 checkerboard where the warped image is overlaid on the fixed inspiratory image. Smooth transitions between the warped and fixed images for axial slices before and after harmonization indicate successful deformation of the moving expiratory image to the fixed inspiratory image.

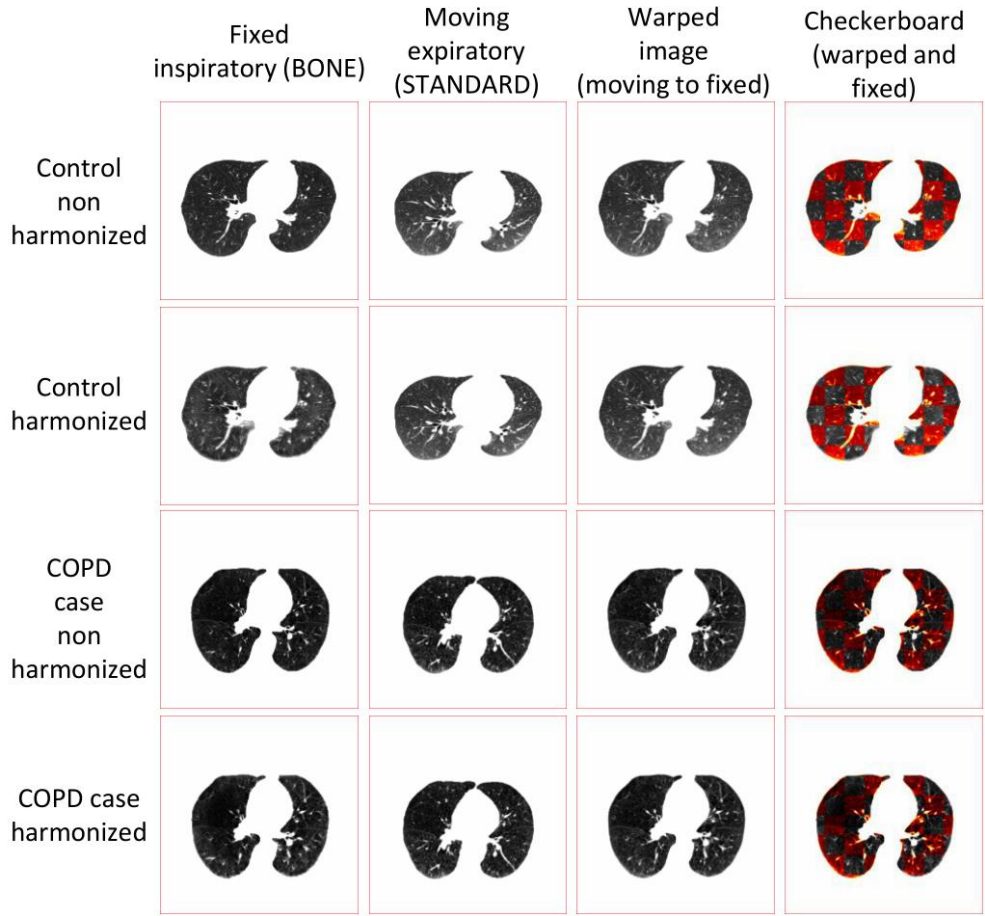

**Figure 4.** Inspiratory scans reconstructed on the BONE kernel sharpen the lung region while expiratory scans reconstructed on a STANDARD kernel is smoother compared to the BONE kernel. Harmonizing the image from a hard BONE kernel to the STANDARD soft kernel standardizes the lung regions to have qualitative consistency. The warped image shows the deformation of the expiratory image to the inspiratory image. We show smooth transitions between the warped image and the fixed image using a checkerboard, highlighting successful registration.

To assess the accuracy of registration both before and after harmonization, we apply the calculated deformation field to warp the emphysema mask of the moving images to the emphysema mask of the fixed image and compare Dice overlap between the fixed and warped emphysema masks. Controls have minimal emphysema as compared to cases. As shown in **Figure 5** (top two rows), before harmonization, the inspiratory BONE kernel scan shows over-segmented emphysema regions as a result of the reconstruction kernel as compared to the expiratory STANDARD kernel scan, which shows minimal emphysema. After harmonization, the over-segmentation on the inspiratory scan is alleviated. The warped emphysema mask shows minimal regions of overlap before and after harmonization. Cases show regions with emphysema as observed by darker regions in the lung CT images. Ideally, an emphysema segmentation algorithm should highlight the same region of the lungs regardless of their reconstruction kernels. However, due to the kernel difference, this is not the case in reality as shown in **Figure 5** (third row). The inspiratory BONE kernel overestimates emphysema as compared to the expiratory STANDARD kernel. However, with harmonization, the overestimation of emphysema on the inspiratory scan is alleviated. The warped emphysema mask (green), overlaid on the fixed inspiratory emphysema mask (red) showcases regions of overlap highlighted in yellow. While the overlap shown by the non-harmonized and harmonized images appear the same, harmonization improves consistency in the emphysema quantification.

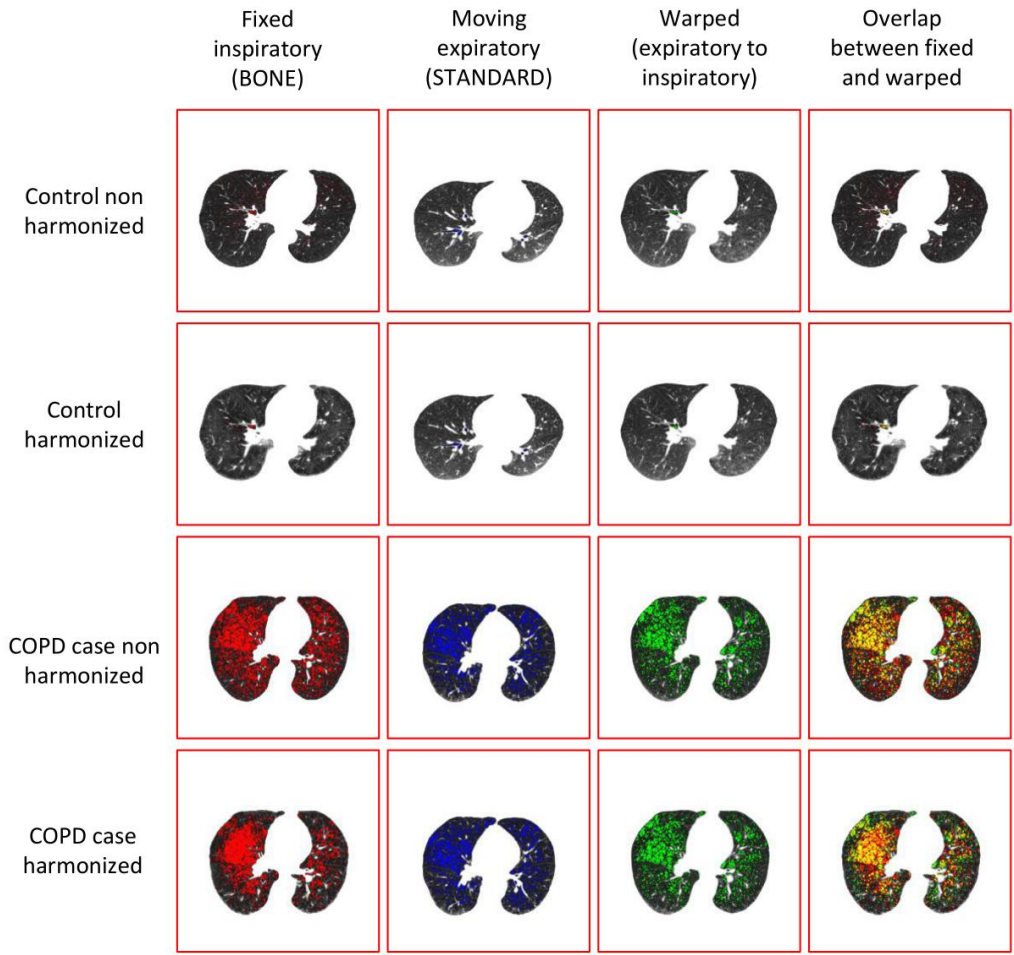

**Figure 5.** We inspect the impact of harmonization and the qualitative accuracy of registration using emphysema for cases and controls in COPDGene. Controls are smokers that have minimal emphysema. The inspiratory BONE kernel scan shows regions of emphysema as compared to the STANDARD kernel expiratory scan as a result of the noise in the image. The harmonized BONE kernel shows minimized emphysema as compared to the non-harmonized kernel. The warped expiratory to inspiratory image shows minimal overlap in emphysema for harmonized and non-harmonized scans. Cases are smokers that exhibit emphysema. The BONE kernel highlights emphysematous regions in harmonized and non-harmonized cases (red), with the non-harmonized case overestimating the measurement. The registered images before and after harmonization show similar overlapping regions (yellow) with the harmonized images showing consistent measurements. The similarity in the overlapping regions between harmonized and unharmonized images indicates that registration does not depend on the kernel.

We visualize the distribution of emphysema scores before and after harmonization using boxplots (**Figure 6**). Before harmonization, the emphysema distribution for the inspiratory scans (BONE) ranges from (1.35, 33.61) with a median emphysema score of 10.48% while the emphysema distribution for the expiratory scans (STANDARD) ranges from (0.06, 21.23) with a median score of 1.41%. After harmonizing the inspiratory scan from the BONE kernel to the STANDARD kernel, the emphysema distribution now ranges from (0.09, 25.32) with a median score of 3.04%. The distribution of emphysema scores gets closer to the target expiratory emphysema distribution, indicating that harmonization reduced the differences in measurements. The differences in measurements before and after harmonization are significantly different as observed from a Wilcoxon signed rank test ($p<0.001$). However, some subjects exhibit higher emphysema scores post harmonization.

To evaluate the performance of deformable registration, we calculate the Dice overlap between the registered and fixed images in three different scenarios after harmonization: a) rigid registration b) deformable image registration at a lower resolution and c) deformable image registration at full resolution. The Dice score distribution for the rigid registered images (rotation and translation) ranges from (0.002, 0.336) with a median score of 0.086. For the deformable registration at half the image resolution, the Dice overlap ranges from (0.005, 0.405) with a median score of 0.125. The deformable registration at full resolution showed improvements in the Dice distribution ranging from (0.014, 0.485) with a median score of 0.148. Paired t-tests between the different groups of registration showed significant differences between them ($p<0.001$), indicating successful registration.

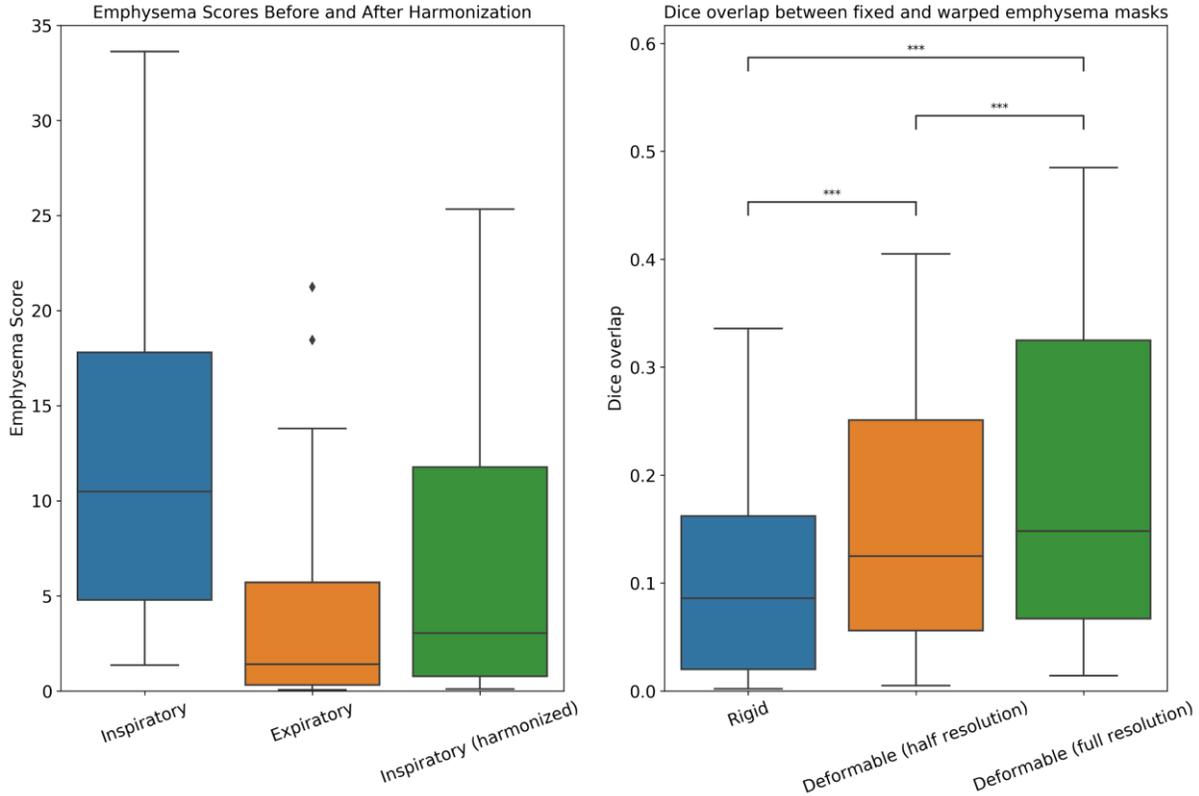

**Figure 6.** The inspiratory BONE kernel overestimates emphysema as compared to the expiratory STANDARD kernel. The STANDARD kernel is preferred for emphysema quantification. Harmonization reduces differences in measurements, enforcing a certain amount of consistency. However, the score distribution indicates that measurements could further be improved. The Dice overlap is computed between the warped emphysema mask from moving to fixed image and the fixed image for the harmonized images. The rigid registration shows low Dice overlap and the Dice overlap improves across the different stages of deformable registration with the full resolution registration showing the highest range of Dice overlap scores.

## 4. DISCUSSION AND CONCLUSION

In this work we investigate the impact of harmonization and registration of paired inspiratory-expiratory scans obtained from the COPDGene study. In the first stage, we harmonize inspiratory scans reconstructed using the BONE kernel to have the style of the expiratory STANDARD kernel using a cycle GAN and quantify emphysema to validate the impact of harmonization. In the second stage we perform deformable image registration by moving the expiratory scans onto the inspiratory scans. We assess the accuracy of registration qualitatively through checkerboards and emphysema overlap between the warped expiratory to inspiratory mask and the fixed inspiratory mask. Additionally, we quantitatively assess Dice overlap between emphysema masks of the warped and fixed image through three different registration stages.

Our harmonization results are consistent with existing findings that investigate CT kernel harmonization. Gallardo-Estrella et al.[30] investigated harmonization on CT kernels obtained from the COPD Gene study. Their approach showed that hard kernels overestimated emphysema and that harmonization showed consistent measurements between the harmonized soft kernel and the ground truth soft kernel. Bak et al.[31] emphasized on emphysema quantification as a validation metric for kernel harmonization and showed that the hard kernel overestimates emphysema three times more than the soft kernel in a COPD cohort. While harmonization does not entirely mitigate differences, we show that measurements become consistent by standardizing the image to the appropriate kernel. In the registration stage, we see that the 3D registration performed by ANTS successfully aligns the moving expiratory image to the fixed inspiratory image through the checkerboard analysis. Moreover, controls show minimal overlap in emphysema regions before and after harmonization while cases show considerable overlap between the emphysema regions of the fixed inspiratory scan and the warped expiratory to inspiratory scan. While Dice overlap obtained on the deformable registration at full resolution is low, we show that there is improvement in the Dice overlap as compared to a rigid registration and deformable registration at half the image resolution.

However, our work is not without its limitations. We use a small number of subjects from the COPD Gene study that comprises of cases and controls. Our study investigated harmonization on reconstruction kernels obtained from the GE manufacturer only. The COPDGene study is a multicenter study that consists of images obtained from multiple scanners, reconstructed using a variety of reconstruction kernels. It is necessary to investigate how different reconstruction kernels impact quantitative imaging. Furthermore, our conclusion on registration is dependent on the ANTs registration algorithm. In future studies, kernel harmonization should be investigated across additional pairs of reconstruction kernels and available registration algorithms need to be evaluated. Through this pilot study, we show that while registration does not depend on harmonization of reconstruction kernels, it is necessary to standardize the image to a reference kernel to obtain accurate quantitative image measurements. Additionally, the deformable registration of inspiratory expiratory scans presents opportunities to quantify volumetric changes through the Jacobian determinant that can provide insights on different categories of COPD.

## ACKNOWLEDGEMENTS

This research was funded by the National Cancer Institute (NCI) grant R01 CA253923. This work was also supported in part by the Integrated Training in Engineering and Diabetes grant number T32 DK101003. This research is also supported by the following awards: National Science Foundation CAREER 1452485; NCI U01 CA196405; UL1 RR024975-01 of the National Center for Research Resources and UL1 TR000445-06 of the National Center for Advancing Translational Sciences; Martineau Innovation Fund grant through the Vanderbilt-Ingram Cancer Center Thoracic Working Group; NCI Early Detection Research Network grant 2U01CA152662. The Vanderbilt Institute for Clinical and Translational Research (VICTR) is funded by the National Center for Advancing Translation Science Award (NCATS), Clinical Translational Science Award (CTSA) Program, Award Number 5UL1TR002243-03. The content is solely the responsibility of the authors and does not necessarily represent the official views of the NIH. We use generative AI to create code segments based on task descriptions, as well as debug, edit, and autocomplete code. Additionally, generative AI technologies have been employed to assist in structuring sentences and performing grammatical checks. It is imperative to highlight that the conceptualization, ideation, and all prompts provided to the AI originate entirely from the authors' creative and intellectual efforts. We take accountability for the review of all content generated by AI in this work.